\documentstyle[psfig,amstex]{europhys}

\def\etal{{\hbox{{\it\ et al.\/}\rm :\ }}}

\def\And{{\rm and\ }}

\def\stars{\bigskip\centerline{***}\medskip}

\newif\ifboo \boofalse

\begin{document}


\def\degre{$^\circ$ }
\newcommand{\microns}{$\mu \text{m}$ }
\newcommand{\ced}{cellule à enclumes en diamant }
\newcommand{\cad}{c.-à-d. }
\newcommand{\poly}{polyèdre }
\newcommand{\tet}{tétraèdre }
\newcommand{\oct}{octaèdre }
\newcommand{\polys}{polyèdres }
\newcommand{\tets}{tétraèdres }
\newcommand{\octs}{octaèdres }
\def\siod{$\text{SiO}_{2}$ }
\def\sioq{$\text{SiO}_{4}$ }
\def\sios{$\text{SiO}_{6}$ }
\def\geod{$\text{GeO}_{2}$ }
\def\geoq{$\text{GeO}_{4}$ }
\def\geos{$\text{GeO}_{6}$ }
\newcommand{\sio}[1]{$\text{SiO}_{#1}$ }
\newcommand{\geo}[1]{$\text{GeO}_{#1}$ }
\newcommand{\alpo}{$\text{AlPO}_{4}$ }
\newcommand{\alaso}{$\text{AlAsO}_{4}$ }
\newcommand{\gapo}{$\text{GaPO}_{4}$ }
\newcommand{\gaaso}{$\text{GaAsO}_{4}$ }
\newcommand{\alo}[1]{$\text{AlO}_{#1}$ }
\newcommand{\po}[1]{$\text{PO}_{#1}$ }
\newcommand{\aso}[1]{$\text{AsO}_{#1}$ }
\newcommand{\gao}[1]{$\text{GaO}_{#1}$ }
\newcommand{\micronsf}{$\mu m$}
\newcommand{\cedf}{cellule à enclumes en diamant}
\newcommand{\polyf}{polyèdre}
\newcommand{\tetf}{tétraèdre}
\newcommand{\octf}{octaèdre}
\newcommand{\polysf}{polyèdres}
\newcommand{\tetsf}{tétraèdres}
\newcommand{\octsf}{octaèdres}
\def\siodf{$\text{SiO}_{2}$}
\def\sioqf{$\text{SiO}_{4}$}
\def\siosf{$\text{SiO}_{6}$}
\def\geodf{$\text{GeO}_{2}$}
\def\geoqf{$\text{GeO}_{4}$}
\def\geosf{$\text{GeO}_{6}$}
\newcommand{\siof}[1]{$\text{SiO}_{#1}$}
\newcommand{\geof}[1]{$\text{GeO}_{#1}$}
\newcommand{\alpof}{$\text{AlPO}_{4}$}
\newcommand{\alasof}{$\text{AlAsO}_{4}$}
\newcommand{\gapof}{$\text{GaPO}_{4}$}
\newcommand{\gaasof}{$\text{GaAsO}_{4}$}
\newcommand{\alof}[1]{$\text{AlO}_{#1}$}
\newcommand{\pof}[1]{$\text{PO}_{#1}$}
\newcommand{\asof}[1]{$\text{AsO}_{#1}$}
\newcommand{\gaof}[1]{$\text{GaO}_{#1}$}

\euro{}{}{}{}

\Date{}

\shorttitle{J. BADRO \etal ON THE HIGH-PRESSURE PHASE TRANSITION ETC.}

\title{On the High-Pressure Phase Transition in GaPO$_{\text{4}}$}

\author{James Badro\inst{1,2}, Jean-Paul Iti\'{e}\inst{2} \And
Alain Polian\inst{2}}

\institute{
     \inst{1} Laboratoire de Sciences de la Terre,
\'{E}cole normale sup\'{e}rieure de Lyon\\
46, all\'{e}e d'Italie 69364 Lyon Cedex 07, France.\\
     \inst{2} Physique des Milieux Condens\'{e}s,
Universit\'{e} Pierre et Marie Curie,  B77\\
 4,  place Jussieu 75252 Paris Cedex 05, France.
}

\rec{}{}

\pacs{
\Pacs{62}{50$+$p}{High Pressure}
\Pacs{61}{50.Ks}{Phase Transitions}
\Pacs{61}{10$-$i}{X-ray diffraction}
      }

\maketitle

\begin{abstract}
X-ray diffraction (XRD) experiments have been carried out on
quartz-like \linebreak
$\alpha -$GaPO$_4$ at high pressure and room temperature. A
transition to a high pressure disordered crystalline form occurs at
13.5~GPa. Slight heating using a YAG infrared laser was applied at
17~GPa in order to crystallize the phase in its stability field. The
structure of this phase is orthorhombic with space group Cmcm. The cell
parameters at the pressure of transition are $a=7.306$~\AA,
$b=5.887$~\AA \ and $c=5.124$~\AA.
\end{abstract}

\section{Introduction}

The low pressure polymorphs of quartz and quartz like compounds are
made up of tetrahedrally coordinated cations. The high
pressure structures consist of octahedral cation arrangements, in which oxygen
atoms are packed in a body-centered cubic (bcc) sub-lattice around the
cations \cite{silica}. At room temperature, the kinetics of these four-
to six-coordinated transitions is very slow leading to a
pressure-induced amorphization (PIA) process observed in quartz
\cite{hemley1, hemley2, meade1, mcneil1, kingma1, silica}. The
compression at room temperature of isostructural compounds $\alpha
-$\geod and $\alpha -$\gaaso has also been studied by X-ray diffraction
and extended X-ray absorption fine structure (EXAFS) spectroscopy
\cite{itie1, badro3, badro4} and the studies show that cation
coordination rises from four to six through the transition.

In the case of berlinites, it is now established that $\alpha -$\alpof,
$\alpha -$\alasof, $\alpha -$\gapo and $\alpha -$\gaaso exhibits a
polymorphic crystalline phase transition, the high pressure form being
very poorly crystallized, and therefore yielding very poor X-ray
diffraction patterns \cite{gillet2, itie3, sowa3, badro4, polian2}.
In-situ EXAFS experiments were realized on $\alpha -$\gapo at the K-edge
absorption energy of
gallium \cite{itie3, polian2} and the polymorphic phase transition
observed at 12~GPa is associated with a four- to six-fold oxygen
coordination change observed around gallium atoms.

\newpage
Molecular dynamics (MD) simulations on berlinite ($\alpha -$\alpof)
associate the transition with a de-stabilization of the \alo{4}
tetrahedron and its transition towards an \alo{6} octahedron
\cite{tse2, watson2}; they also have concluded that room temperature
PIA would occur only when the \po{4} tetrahedra went unstable, at a
predicted pressure of 80~GPa, this pressure having not yet been produced
experimentally on \alpof.

\section{Experimental}

Quartz structured gallium phosphate single crystals grown by the
hydrothermal technique were crushed into powder and used for this high
pressure energy dispersive X-ray diffraction study. Platinum powder was
mixed with the sample in order to absorb the infrared beam emmited by a
YAG laser and heat the system. The mixture was then
loaded in a stainless steel gasket with a 250 $\mu \text{m}$ hole
diameter and 40 $\mu \text{m}$ initial thickness and high pressure
was generated by a membrane driven \cite{letoullec1} diamond anvil cell (DAC).
Nitrogen was used as pressure transmitting medium, and ruby
fluorescence as well as platinum isothermal equation of state (EOS)
were used as pressure gauges; The measurements were carried out on the
DW-11 energy dispersive X-ray diffraction beamline on the DCI ring of
the LURE synchrotron facility in Orsay, France. Accumulation times
ranged from 1~hour at low pressure and up to 12 hours in the high
pressure phase with a diffraction angle $2\theta=12^\circ$. At 17~GPa,
the sample was heated by diffusion of heat from the infrared absorbing
platinum mixed with the sample.

\section{Compression and Heating}

The high-pressure behaviour of the quartz phase of \gapo has been
studied by X-ray diffraction \cite{sowa2, polian2, itie3}, Raman
\cite{wolfunp, polian2} and Brillouin \cite{polian2} scattering.

Isothermal equations of state (EOS) have been fitted on pressure-volume
data and yield similar results. As with other berlinite phases, it was
assumed that  \gapo transformed to an amorphous phase \cite{sowa2} at
9~GPa, but new experiments \cite{polian2, itie3} show that this
transformation is instead due to a polymorphic crystalline transition.

\begin{figure}[h!]
\centerline{\psfig{figure=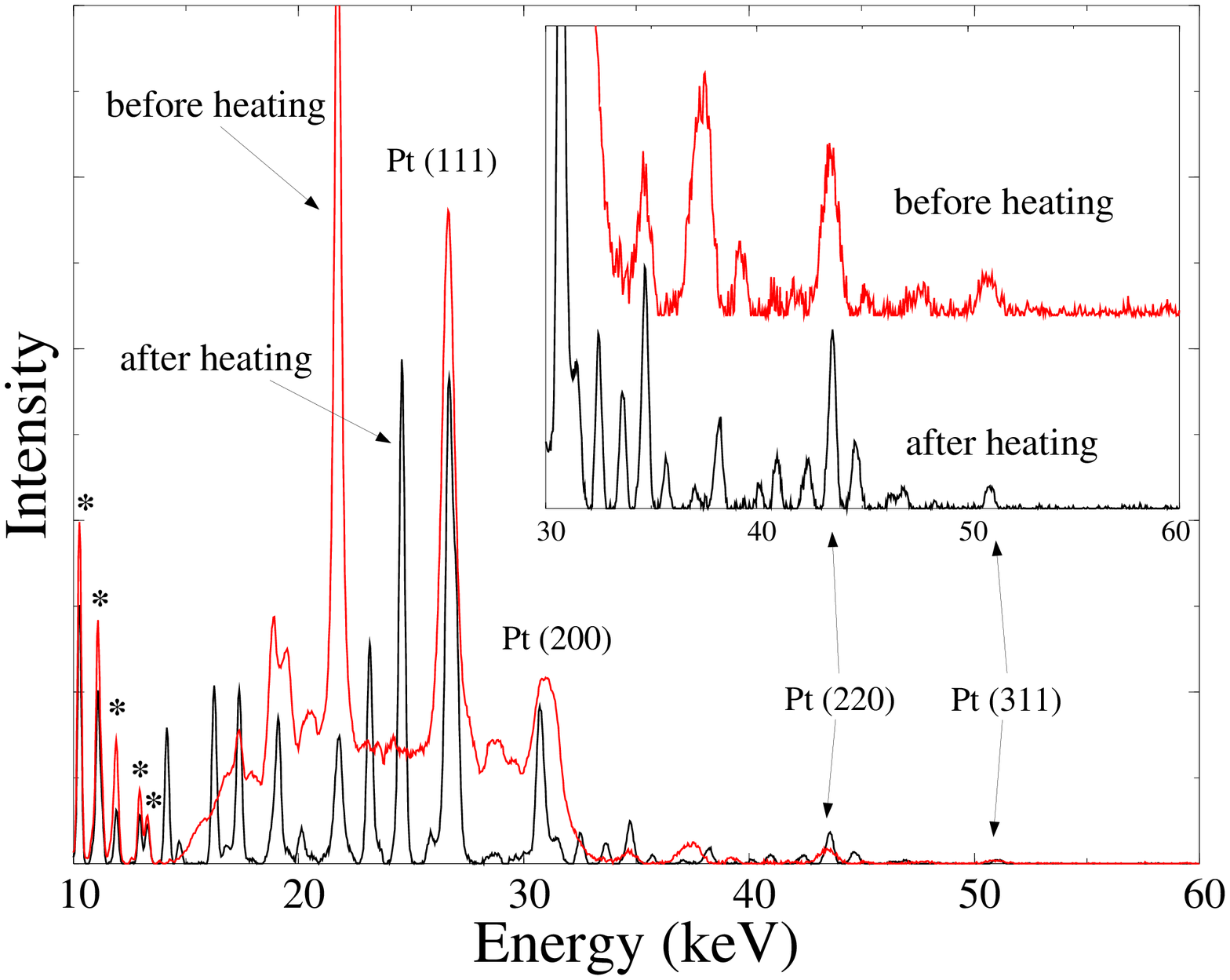,width=13cm}}
\caption{X-ray diffraction spectra of \gapo at 17~GPa
before and after laser heating. The heated sample exhibits a finer
crystalline state, and care was taken in order to ensure no phase
transition occured upon heating. Gallium and platinum fluorescence
lines are marked with a star (*) sign (in the 10--13~keV range).
Platinum diffraction peaks are also indicated on the
figure, along with the corresponding Miller indices. The inset shows a
detailed view of the spectra in the high energy domain and we observe
that most of the "new" diffraction lines were already present, but were
broad and weak.}
\label{fig:edx_gapo4}
\end{figure}

In this experiment, very few points were produced at low pressure. At a
pressure of 17~GPa, the sample had already crossed the transition
pressure and the spectrum of the high-pressure phase is reproduced in
figure~\ref{fig:edx_gapo4}. It was impossible to index the peaks due to
both the disorder of the phase and the broadening induced by the very
small size of the crystallites in the high pressure phase after such
thermally inhibited reconstructive phase transitions. At this point,
the sample was heated
in order to "activate" the kinetics of the transition. Care was taken
in order to insure that no phase transition occurred during heating. The
system was brought back to ambient temperature and a diffraction
spectrum was collected. The sample was very well crystallized and the
diffraction lines (figure~\ref{fig:edx_gapo4}) can be indexed in an
orthorhombic system (table~\ref{table:dhkl_gapo4}) with space group
$\text{C}_{\text{mcm}}$, and the phase is of $\text{InPO}_{4}$
structure-type, with octahedral gallium and tetrahedral phosphorus
atoms \cite{itie3, badro4, polian2}.
At this pressure, the unit cell parameters were
$a=7.306(7)$~\AA, $b=5.887(6)$~\AA \ and
$c=5.124(3)$~\AA. A detailed analysis of these two spectra
shows that no transition occurred during heating, because none of the
existing diffraction lines before heating had vanished afterwards, and
most of the new line which appeared after heating were already present
before but were broad and very weak (figure~\ref{fig:edx_gapo4}). It is only
after such an analysis that we could infer that these two spectra are
that of the same thermodynamic phase, one of which is much better
crystallized than the other. It can be noted here that this is the
first characterization, though indirect, of the high-pressure and
room-temperature phase of $\alpha -$\gapof.

It is interesting to note that this phase is similar to the one
obtained by room temperature compression of a high-temperature
polymorph of \gapof, namely the low cristobalite phase \cite{robeson1,
murashov1}. This finding confirms that the transition of the quartz
phase at high pressure is thermally inhibited because cristobalite
being metastable at room temperature, it has a higher free enthalpy
than the quartz-form and therefore it necessarily has a lower thermodynamic
energy barrier with respect to the high pressure orthorhombic phase, as
well as a higher driving force for the transition. It can therefore
transit to this structure without the need for kinetic energy
activation brought by heating.

\begin{figure}[h]
\hspace*{2cm}{\psfig{figure=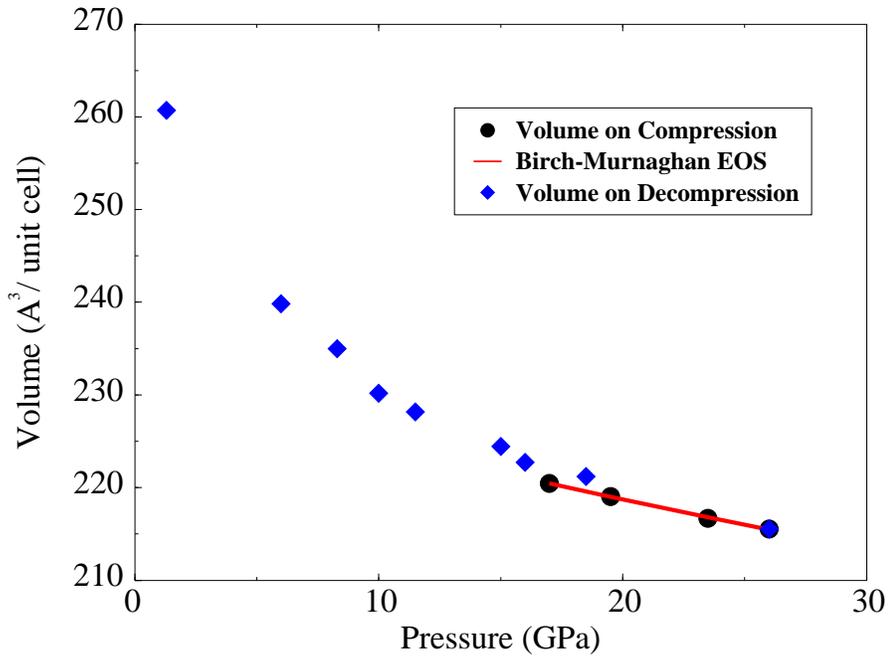,width=10cm}}
\caption{Pressure--Volume data points obtained from
refinement of the orthorhombic structure of the sample in the high
pressure phase. The points are fitted by a Birch-Murnaghan third order
EOS. Observed values for the variables are: $\text K_{0}=$~308~GPa and
$\text V_{0}=$~231.7~\AA$^{3}$, with the pressure derivative of the
bulk modulus set to 4. This equation is fitted with four points (filled
circles) from 17~GPa up to 26~GPa. The points on decompression (filled
diamonds) were not used, because the apparition of deviatoric and
uniaxial stresses in the sample makes volume data unreliable (for
example, point at 18.5~GPa on decompression).}
\label{fig:eos_gapo4}
\end{figure}

The system was then further compressed in order to measure the bulk
modulus of the new high pressure phase. Three points at higher pressure
were undertaken, and a Birch-Murnaghan third order equation of state
was used for fitting through the points (figure~\ref{fig:eos_gapo4}). The
pressure derivative of the bulk modulus was fixed at $\text
K^{'}_{0}=$~4, the unit cell volume at ambient pressure was found to be
$\text V_{0}=$~231.7~\AA$^{3}$ and the fitted bulk modulus equal to
$\text K_{0}=$~308~GPa.

\section{Decompression}

A step by step decompression path was then followed in order to have a
precise view on the back-transformation of the sample. The high
pressure phase is retained down to 6~GPa. At 1~GPa, the (200) peak of
$\alpha -$\gapo along with other weak peaks appear in the diffraction
pattern, together with the peaks of the high pressure orthorhombic phase.
Thus, the quenched sample consists of a mixture of both phases, meaning the
the transition is not fully reversible. It should be noted that these
points obtained on decompression that were still in the high pressure
phase were not included in the fit for the EOS, because a huge amount
of stress appears at downstroke, and the points do not generally
coincide with those obtained at compression as can be seen at 18.5~GPa
in figure~\ref{fig:eos_gapo4}.

\enlargethispage{5cm}

\section{Conclusion}

High-pressure heating experiments have been undertaken and show that an
increase in the temperature enhances dramatically the crystallization
of the sample in the high-pressure phase, thus yielding narrower
peaks.  The phase was then indexed in an orthorhombic symmetry in the
Cmcm space group. The high pressure phase is of $\text{InPO}_{4}$ type,
with octahedral gallium and tetrahedral phosphorus.  The high pressure
phase obtained seems to be identical to the high pressure crystal
formed by the room temperature compression of the high temperature
polymorph. This seems to confirm the fact that the transition from the
P3$_1$21 quartz-like material to the Cmcm, $\text{InPO}_{4}$-like phase
is thermally inhibited at room temperature. Even though this transition
does take place in the stability field of the phase, room temperature
kinetics are not sufficient to drive the transition. This confirms
that the high pressure phase transition in \gapo is a polymorphic
crystalline phase transition and not a pressure-induced amorphization,
as previously reported \cite{sowa2}.
The observed gallium four-fold to six-fold coordination change at room
temperature is therefore associated with the expected
quartz--to--InPO$_4$ transformation.

A similar transformation occurs in $\alpha-$\alpo at room temperature,
at least from the spectroscopic point of view, and these two materials,
{\em i.e.\/} \alpo and \gapof, exhibit a similar high-pressure
behaviour.
It has very recently been shown that isostructural FePO$_4$ undergoes a
similar transformation to the InPO$_4$ structure \cite{pasternak97} at high
pressure.
This study shows that it is therefore likely that the
InPO$_4$ phase is the host structure for phosphate berlinites at high
pressure.

In this transformation, the phosphorus tetrahedra remain unaltered.  At
room-temperature, the partial back-transformation to the quartz-phase
is a mirror of the high-pressure transformation. The
back-transformation associated with a conservation of the
crystallographic axes is a direct consequence of the rigidity of these
\po{4} clusters.

\stars

We wish to express our grateful thanks to our colleagues Fabrice
Visocekas and Denis Andrault of the Département des Géomatériaux
at the Institut de Physique du Globe de Paris for carrying out
the laser heating of the sample. We sincerely acknowledge Philippe Gillet
for his comments and discussions on this work.

Laboratoire de Sciences de la Terre is CNRS UMR 5570.  Physique des
Milieux Condens\'{e}s is CNRS URA 782.

\newpage

\newpage

\begin{table}[h] 
\caption{\label{table:dhkl_gapo4} Observed and calculated (hkl)
spacings for the high pressure phase at 17~GPa after heating. The
calculated lines are that of an orthorhombic system with a centered
Bravais lattice. The cell parameters are $\text a=7.306$~\AA, $\text
b=5.887$~\AA \ and $\text c=5.124$~\AA. The difference between observed
and calculated spacings is taken using 5 decimal places for observed
lines, but the latter can reasonable de given with only three
significant digits.}
\vspace{6mm}
\centerline{
\begin{tabular}{|c|c|c|r||c|c|c|r|} \hline
{\rule[-3mm]{0mm}{6mm} \raisebox{-0.8ex}{$(hkl)$}}
& \raisebox{-0.8ex}{$d_{obs}$
(\AA)} & \raisebox{-0.8ex}{$d_{cal}$ (\AA)} &
\raisebox{-0.8ex}{$d_{obs}$-$d_{cal}$ (\AA)} &
\raisebox{-0.8ex}{$(hkl)$} & \raisebox{-0.8ex}{$d_{obs}$
(\AA)} & \raisebox{-0.8ex}{$d_{cal}$ (\AA)} &
\raisebox{-0.8ex}{$d_{obs}$-$d_{cal}$ (\AA)}\\ \hline
     $ 1 0 1$    &    4.195 &  4.19543 & -0.00026 \hspace*{4pt}&
     $ 2 0 0$    &    3.650 &  3.65313 & -0.00360 \hspace*{4pt}\\
     $ 1 1 1$    &    3.419 &  3.41677 &  0.00211 \hspace*{4pt}&
     $ 2 1 0$    &    3.107 &  3.10419 &  0.00272 \hspace*{4pt}\\
     $ 0 2 0$    &    2.946 &  2.94397 &  0.00207 \hspace*{4pt}&
     $ 1 2 0$    &    2.725 &  2.73063 & -0.00601 \hspace*{4pt}\\
     $ 2 1 1$    &    2.661 &  2.65506 &  0.00567 \hspace*{4pt}&
     $ 0 0 2$    &    2.562 &  2.56226 & -0.00064 \hspace*{4pt}\\
     $ 1 0 2$    &    2.414 &  2.41789 & -0.00402 \hspace*{4pt}&
     $ 1 2 1$    &    2.413 &  2.40986 &  0.00401 \hspace*{4pt}\\
     $ 3 0 1$    &    2.198 &  2.19965 & -0.00160 \hspace*{4pt}&
     $ 1 3 0$    &    1.888 &  1.89545 & -0.00717 \hspace*{4pt}\\
     $ 4 0 0$    &    1.825 &  1.82657 & -0.00169 \hspace*{4pt}&
     $ 3 0 2$    &    1.764 &  1.76524 & -0.00149 \hspace*{4pt}\\
     $ 0 0 3$    &    1.710 &  1.70817 &  0.00186 \hspace*{4pt}&
     $ 1 0 3$    &    1.662 &  1.66332 & -0.00109 \hspace*{4pt}\\
     $ 1 1 3$    &    1.600 &  1.60068 & -0.00091 \hspace*{4pt}&
     $ 0 2 3$    &    1.479 &  1.47748 &  0.00138 \hspace*{4pt}\\
     $ 1 2 3$    &    1.449 &  1.44816 &  0.00094 \hspace*{4pt}&
     $ 3 0 3$    &    1.400 &  1.39848 &  0.00166 \hspace*{4pt}\\
     $ 4 2 2$    &    1.328 &  1.32753 &  0.00030 \hspace*{4pt}&
 & & & \\
\hline
\end{tabular}
}
\end{table}

\newpage
\vspace*{2cm}

\end{document}